\documentclass[twocolumn,superscriptaddress]{revtex4-1}
\usepackage{bm,color}
\usepackage[dvipdfmx]{graphicx}
\usepackage{amsmath}

\begin{document}
\title {Dynamical phase transitions in the photodriven charge-ordered Dirac-electron system}

\author{Yasuhiro Tanaka}
\affiliation{Department of Applied Physics, Waseda University, Okubo, Shinjuku-ku, Tokyo 169-8555, Japan}
\author{Masahito Mochizuki}
\affiliation{Department of Applied Physics, Waseda University, Okubo, Shinjuku-ku, Tokyo 169-8555, Japan}
\begin{abstract}
Photoinduced phase transitions and charge dynamics in the interacting Dirac-electron system with a charge-ordered ground state are theoretically studied by taking an organic salt $\alpha$-(BEDT-TTF)$_2$I$_3$. By analysing the extended Hubbard model for this compound using a combined method of numerical simulations based on the time-dependent Schr\"odinger equation and the Floquet theory, we observe successive dynamical phase transitions from the charge-ordered insulator to a gapless Dirac semimetal and, eventually, to a Chern insulator phase under irradiation with circularly polarized light. These phase transitions occur as a consequence of two major effects of circularly polarized light, i.e., closing of the charge gap through melting the charge order and opening of the topological gap by breaking the time reversal symmetry at the Dirac points. We demonstrate that these photoinduced phenomena are governed by charge dynamics of driven correlated Dirac electrons.
\end{abstract}
\maketitle

\begin{figure}[tb]
\includegraphics[scale=0.5]{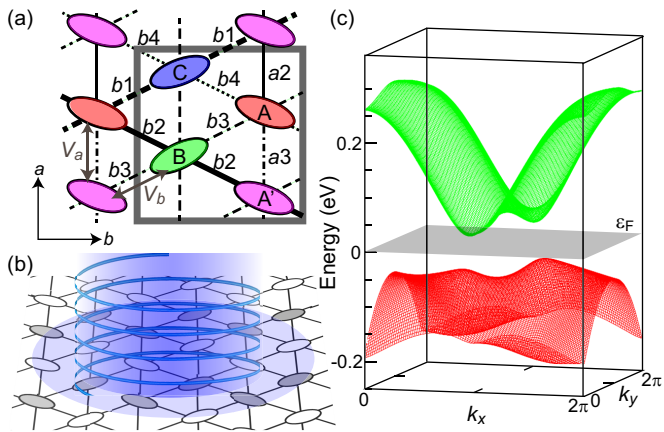}
\caption{(a) Conduction layer of $\alpha$-(BEDT-TTF)$_2$I$_3$. The ellipses represent the BEDT-TTF molecules, and the gray rectangle indicates the unit cell which contains four molecules (A, A$^{\prime}$, B, and C). In the charge-ordered phase, the sites A$^{\prime}$ and C are electron-rich, whereas the sites A and B are electron-poor. (b) Schematic illustration of the photoirradiation on charge-ordered $\alpha$-(BEDT-TTF)$_2$I$_3$ with circularly polarized light. The shaded and open ellipses represent the charge disproportionation. (c) Dispersion relations of bands above and below the Fermi level for the charge-ordered phase.}
\label{Fig01}
\end{figure}
Photoirradiation causes dramatic changes of electronic and magnetic properties in solids~\cite{Nasu_Book,Kirilyuk_RMP10}. A typical example is the photoinduced insulator-to-metal transitions in strongly correlated electron systems where charge-localized  insulating states caused by Coulomb interactions melt under application of intense laser pulse~\cite{Yonemitsu_PR08,Ishihara_JPSJ19,Okamoto_PRB10,Iwai_PRL07}. Another intriguing phenomenon is photoinduced topological phase transitions in the noninteracting Dirac-electron systems where irradiation with circularly polarized light induces a topological gap at the Dirac points through breaking the time reversal symmetry and renders the system topologically nontrivial Chern insulator~\cite{Haldane_PRL88,Oka_PRB09,Kitagawa_PRB10,Bukov_AP15,Mclver_NP20,Sato_PRB19,Schuler_PRX20}.

When the electron-electron interactions manifest themselves in the Dirac semimetals, they can induce an order of charge degrees of freedom, e.g., Mott insulating order and charge order, which opens a charge gap at the Dirac points and renders the systems nontopological insulator~\cite{Sorella_EPL92,Herbut_PRL06,Raghu_PRL08}. Then a question arises, that is, what happens when this interacting Dirac-electron system is irradiated by circularly polarized light? The photoirradiation should work to close the charge gap through melting the charge order, but it could simultaneously work to open a topological gap at the Dirac points. Interplay and competition between these two dynamical effects of light are nontrivial, and how these two different dynamical phase transitions appear in the photodriven interacting Dirac electrons is an issue of interest.

The organic conductor $\alpha$-(BEDT-TTF)$_2$I$_3$ provides us a unique opportunity to study this problem. This compound has quasi-two-dimensional conduction layers composed of four BEDT-TTF molecules (A, A$^\prime$, B, and C) in the unit cell, which form an anisotropic triangular lattice [Fig.~\ref{Fig01}(a)]~\cite{Mori_CL84,Kakiuchi_JPSJ07}. At ambient pressure, this material exhibits a horizontal-stripe charge order below 135 K~\cite{Takano_JPCS01,Woj_PRB03}, stabilized by the long-range Coulomb interactions~\cite{Kino_JPSJ95,Seo_JPSJ00,Tanaka_JPSJ08,Tanaka_JPSJ16}. This charge order vanishes when a uniaxial pressure $P_a(>4\,{\rm kbar})$ is applied, and the system becomes the Dirac semimetal with a pair of gapless Dirac-cone bands whose Dirac points are located on the Fermi level~\cite{Katayama_JPSJ06,Tajima_JPSJ06}. These facts indicate that this material is a strongly correlated Dirac-fermion system. In fact, the melting of charge order and the formation of metallic domains at ambient pressure were experimentally observed under irradiation of {\it linearly} polarized light by pump-probe measurements~\cite{Iwai_PRL07}. On the other hand, possible photoinduced topological phase transitions under irradiation with {\it circularly} polarized light have been theoretical proposed using a (noninteracting) tight-binding model for the Dirac semimetal phase under pressures~\cite{Kitayama_PRR20,Kitayama_JPSJ21,Tanaka_PRB21}. However, the effects of photoirradiation with circularly polarized light on the charge-ordered phase have been studied neither experimentally nor theoretically so far, despite novel fundamental physics of photodriven correlated Dirac fermions are anticipated.

In this Letter, we theoretically study the dynamical phase transitions and charge dynamics induced by irradiation with circularly polarized light in the organic salt $\alpha$-(BEDT-TTF)$_2$I$_3$ as a typical example material of the correlated Dirac-electron systems with a charge-ordered ground state. By analysing an extended Hubbard model for this compound using a combined method of the numerical simulation based on the time-dependent Schr\"odinger equation and the Floquet theory, we reveal that successive photoinduced phase transitions from the charge order to a gapless Dirac semimetal to a Floquet Chern insulator occur accompanied by closing of the insulating gap of charge-order origin and subsequent opening of the topological gap due to the broken time-reversal symmetry at the Dirac points. It is revealed that the photodriven charge dynamics governs these novel photoinduced phenomena in the interacting Dirac-electron system.

We start with an extended Hubbard model for the BEDT-TTF layer in $\alpha$-(BEDT-TTF)$_2$I$_3$~\cite{Kino_JPSJ95,Seo_JPSJ00,Tanaka_JPSJ08,Tanaka_JPSJ10,Miyashita_JPSJ10},
\begin{eqnarray}
{\mathcal H}(\tau)&=&\sum_{\langle i,j\rangle, \sigma}(t_{ij}e^{i(e/\hbar){\bm \delta}_{ij}\cdot {\bm A}(\tau)}
c^{\dagger}_{i\sigma}c_{j\sigma}+{\rm H.c.}) \nonumber \\
&+&U\sum_{i}n_{i\uparrow}n_{i\downarrow}+\sum_{\langle i,j\rangle}V_{ij}n_in_j.
\label{eq1}
\end{eqnarray}
Here $\langle i,j\rangle$ represents the summation over pairs of adjacent molecular sites, $c^{\dagger}_{i\sigma}$ ($c_{i\sigma}$) is the creation (annihilation) operator for an electron with spin $\sigma$ at site $i$, $n_{i\sigma}=c^{\dagger}_{i\sigma}c_{i\sigma}$, and $n_i=n_{i\uparrow}+n_{i\downarrow}$. The transfer integrals $t_{ij}$ are given by seven parameters $t_\ell$ with $\ell$ being the bond index shown in Fig.~\ref{Fig01}(a), i.e., $t_{b1}=0.127$ eV, $t_{b2}=0.145$ eV, $t_{b3}=0.062$ eV, $t_{b4}=0.025$ eV, $t_{a1}=-0.035$ eV, $t_{a2}=-0.046$ eV, and $t_{a3}=0.018$ eV~\cite{Kakiuchi_JPSJ07}. For the repulsive Coulomb interactions $U$ and $V_{ij}$, we take $U=0.6$ eV, $V_a=0.42U$, and $V_b=0.6V_a$ where $V_a$ ($V_b$) is the interactions between the sites connected by vertical (diagonal) bonds [Fig. \ref{Fig01}(a)].

The four molecules in the unit cell compose four bands around the Fermi level. Because the electron filling is 3/4, the Fermi level is located between the third and fourth bands. Figure~\ref{Fig01}(c) presents dispersion relations of these two bands for the mean-field solution of the horizontal charge-ordered state before photoirradiation, in which a charge gap is opened at the Dirac points. In this state, the sites A$^\prime$ and C are electron-rich, whereas the sites A and B are electron-poor. Consequently, the charge disproportionation $\langle n_{\rm A^\prime}\rangle-\langle n_{\rm A}\rangle$ is positively finite, but it vanishes when the charge order disappears. Therefore, this quantity can be exploited as an order parameter of this charge order. On the contrary, the relation $\langle n_{\rm B}\rangle<\langle n_{\rm C}\rangle$ holds even in the absence of charge order because of the crystal symmetry~\cite{Kobayashi_JPSJ04}.

The effects of photoirradiation are considered through the Peierls phases attached to the transfer integrals. We define $\bm \delta_{ij}=\bm r_j-\bm r_i$ where $\bm r_i$ is the positional vector of site $i$. We examine a pulse of circularly polarized light [Fig.~\ref{Fig01}(b)]. The vector potential is given by,
\begin{eqnarray}
\bm A(\tau)=\frac{E^\omega}{\omega}\exp\left[-\frac{(\tau-\tau_{\rm pu})^2}{2\sigma_{\rm pu}^2}\right](\cos \omega\tau, \sin \omega\tau),
\label{eq2}
\end{eqnarray}
where $E^{\omega}$ and $\omega$ are the amplitude and frequency of light electric field, respectively. The center and width of the pulse are set as $\tau_{\rm pu}=250T$ and $\sigma_{\rm pu}=76T$, respectively, while the frequency of light is fixed at $\hbar\omega=0.7$ eV. Hereafter we use natural units with $e=\hbar=1$, and the lattice constant is chosen as the unit of length~\cite{Tanaka_PRB21}.

We simulate real-time charge dynamics by numerically integrating the time-dependent Schr\"odinger equation after applying the mean-field approximation to the Hamiltonian in Eq.~(\ref{eq1}),
\begin{equation}
|\psi_{{\bm k},\nu}(\tau+d\tau)\rangle={\mathcal T}{\rm exp}\Bigl[-i\int^{\tau+d\tau}_{\tau}
d\tau^{\prime}{\mathcal H}^{\rm MF}_{{\bm k}}(\tau^{\prime})\Bigr]|\psi_{{\bm k},\nu}(\tau)\rangle.
\label{eq3}
\end{equation}
Here ${\mathcal H}^{\rm MF}_{{\bm k}}(\tau)$ is a $4\times 4$ matrix of the mean-field Hamiltonian in the momentum space, $|\psi_{{\bm k},\nu}(\tau)\rangle$ is the $\nu$th ($\nu$=1-4) one-particle state with wavevector ${\bm k}$ at time $\tau$, and ${\mathcal T}$ denotes the time-ordering operator.  The spin degrees of freedom are not incorporated in Eq.~(\ref{eq3}) because it is known that the charge-ordered phase in this compound is nonmagnetic~\cite{Rothaemel_PRB86}. We set $d\tau=T/M$ with $M=800$ where $T=2\pi/\omega$ is the period of light, and expand the exponential operator in Eq.~(\ref{eq3}) with $d\tau$ such that the wavefunctions $|\psi_{{\bm k},\nu}(\tau)\rangle$ are calculated within an error of the order of $(d\tau)^3$~\cite{Kuwabara_JPSJ95,Terai_PTPS93,Tanaka_JPSJ10}. A system of $100 \times 100$ unit cells is used for the simulations.

\begin{figure}[tb]
\includegraphics[scale=1.0]{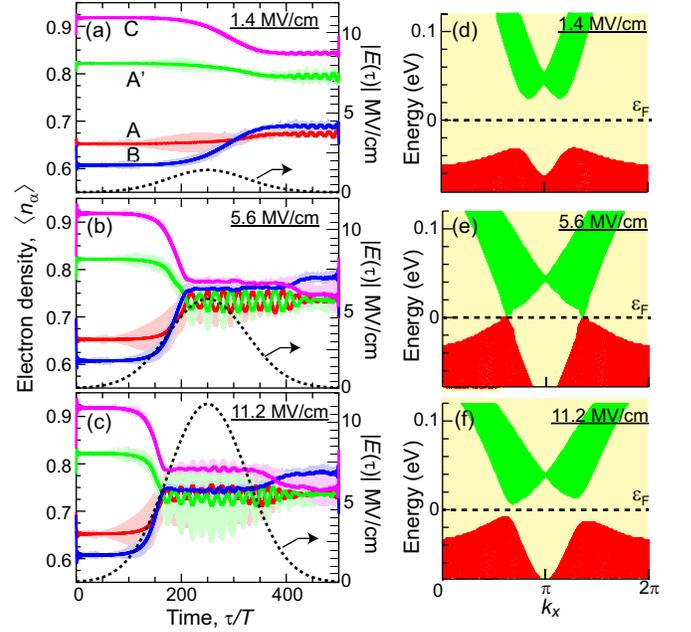}
\caption{(a)-(c) Time evolutions of electron densities $\langle n_{\alpha}\rangle$ at four molecular sites ($\alpha=$A, A$^\prime$, B, and C) for various light intensities, i.e., (a) $E^\omega$=1.4 MV/cm, (b) $E^\omega$=5.6 MV/cm, and (c) $E^\omega$=11.2 MV/cm, when the light frequency is $\hbar\omega$=0.7 eV. The bold solid lines present the results after the high-frequency components are filtered out. Time profiles of the light amplitude $|\bm E(\tau)|$ are also depicted by dashed lines. (d)-(f) Corresponding projected band dispersions of the transient states.}
\label{Fig02}
\end{figure}
Figures~\ref{Fig02}(a)-\ref{Fig02}(c) show simulated time evolutions of electron densities $\langle n_{\alpha}\rangle$ at the four molecular sites ($\alpha=$A, A$^{\prime}$, B, C) for various strengths of light electric field, i.e., (a) weak light field of $E^\omega$=1.4 MV/cm, (b) intermediate light field of $E^\omega$=5.6 MV/cm, and (c) strong light field of $E^\omega$=11.2 MV/cm. The electron densities $\langle n_{\alpha}\rangle$ exhibit fast oscillations with frequencies comparable to the light frequency $\omega$. Therefore, we extract slow-oscillation components through filtering out the components faster than $0.5\omega$.

When the light field is as weak as $E^\omega$=1.4 MV/cm, the charge order with $\langle n_{\rm A^\prime}\rangle-\langle n_{\rm A}\rangle>0$ survives even after the photoirradiation [Fig. \ref{Fig02}(a)]. On the contrary, for the intermediate light field of $E^\omega$=5.6 MV/cm, the charge order vanishes, i.e., $\langle n_{\rm A^\prime}\rangle-\langle n_{\rm A}\rangle \sim 0$ at $\tau/T\sim 200$ [Fig. \ref{Fig02}(b)]. After the melting of charge order ($\tau/T> 200$), the electron densities $\langle n_{\rm A}\rangle$ and $\langle n_{\rm A^\prime}\rangle$ slowly oscillate around the same oscillation centers ($\overline{\langle n_{\rm A}\rangle} = \overline{\langle n_{\rm A^\prime}\rangle}\sim$0.73) with a time period of $\sim 30T$, indicating that the A and A$^\prime$ sites become equivalent. This also means that static components of the Hartree potentials at sites A and A$^\prime$, $\phi_{\rm A}$ and $\phi_{\rm A^\prime}$, are equivalent, which are given by $\phi_{\rm A}=2V_a\langle n_{\rm A^\prime}\rangle+2V_b(\langle n_{\rm B}\rangle +\langle n_{\rm C}\rangle)$ and $\phi_{\rm A^\prime}=2V_a\langle n_{\rm A}\rangle+2V_b(\langle n_{\rm B}\rangle +\langle n_{\rm C}\rangle)$, respectively.

When the light field is as strong as $E^\omega$=11.2 MV/cm, the melting of charge order occurs more quickly at $\tau/T\sim 150$. Around the pulse center at $\tau/T=250$, the electron densities $\langle n_{\rm A}\rangle$ and $\langle n_{\rm A^\prime}\rangle$ show prominent fast oscillations, while their slow-oscillation components are suppressed. Notably, centers of the slow oscillations differ between $\langle n_{\rm A}\rangle$ and $\langle n_{\rm A^\prime}\rangle$, indicating inequivalence of the sites A and A$^\prime$ (or the site-potentials $\phi_{\rm A}$ and $\phi_{\rm A^\prime}$).

We then investigate how this photoinduced charge dynamics affects the nonequilibrium electronic structure by calculating the excitation spectra in the transient processes~\cite{Freericks_PRL09,Sentef_NatCom15},
\begin{eqnarray}
A_{\bm k}(\varepsilon, \tau_{\rm pr})={\rm Im}\sum_{\alpha}\int d\tau_1 d\tau_2
s(\tau_1-\tau_{\rm pr})s(\tau_2-\tau_{\rm pr}) \nonumber \\
e^{i\varepsilon (\tau_1-\tau_2)}[G^{<}_{{\bm k},\alpha\alpha}(\tau_1,\tau_2)
-G^{>}_{{\bm k},\alpha\alpha}(\tau_1,\tau_2)],
\end{eqnarray}
with
\begin{align}
&G^{<}_{{\bm k},\alpha\beta}(\tau_1,\tau_2)=i\langle c^{\dagger}_{{\bm k},\beta}
(\tau_2)c_{{\bm k},\alpha}(\tau_1)\rangle,
\\
&G^{>}_{{\bm k},\alpha\beta}(\tau_1,\tau_2)=
-i\langle c_{{\bm k},\alpha}(\tau_1)c^{\dagger}_{{\bm k},\beta}(\tau_2)\rangle,
\end{align}
where 
$G^{<}_{{\bm k},\alpha\beta}$ and $G^{>}_{{\bm k},\alpha\beta}$ are the lesser and greater Green's functions, respectively. The operator $c^{\dagger}_{{\bm k},\alpha}$ and $c_{{\bm k},\alpha}$ are the Fourier transforms of $c^{\dagger}_{\gamma, \alpha}$ and $c_{\gamma, \alpha}$ where $\gamma$ is the index of unit cells. We assume the following Gaussian function for a probe pulse,
\begin{eqnarray}
s(\tau-\tau_{\rm pr})=\frac{1}{\sigma_{\rm pr}\sqrt{2\pi}}
\exp \left[-\frac{(\tau-\tau_{\rm pr})^2}{2\sigma_{\rm pr}^2}\right],
\end{eqnarray}
where the center and width of the pulse are set as $\tau_{\rm pr}=\tau_{\rm pu}$ and $\sigma_{\rm pr}/T=25$. 

Figures~\ref{Fig02}(d)-(f) show band structures in the transient states around the moment corresponding to the pulse center, which are determined by the peak positions of $A_{\bm k}(\varepsilon,\tau_{\rm pr})$ projected onto the ($k_x$, $\varepsilon$)-plane. Two gaps at the Dirac points are denoted by $\Delta_1$ and $\Delta_2$ for $k_x<\pi$ and for $k_x>\pi$, respectively. For the weak light field of $E^\omega$=1.4 MV/cm, the band structure has a large insulating gap in the presence of charge order that survives against the photoexcitation. When the light field is intermediate in strength ($E^\omega$=5.6 MV/cm), the charge order melts, which results in the gap closing ($\Delta_1=\Delta_2=0$) and the emergence of photoinduced Dirac semimetal phase. For the strong light field of $E^\omega$=11.2 MV/cm, the gap opens again which indicates the emergence of nonequilibrium topological insulator called Floquet Chern insulator~\cite{Tanaka_PRB21}. 

To investigate the topological nature of the photoinduced phases, we need to calculate the Berry curvatures and the Chern numbers of these pseudo nonequilibrium steady phases. These quantities for photodriven systems are usually calculated by the Floquet Hamiltonian. However, the Floquet theory cannot be applied to the driven interacting systems directly because the time-dependence of the Hamiltonian including the dynamical two-body interaction terms is not trivial. In this work, we construct a Floquent Hamiltonian for the interacting Dirac-electron system by considering the simulated time profiles of mean-field order parameters $\rho_{ij}(\tau) \equiv \sum_{\sigma}\langle c^{\dagger}_{i\sigma}c_{j\sigma}\rangle$. The Fourier transforms of $\rho_{ij}(\tau)$ are performed within a time domain around the pulse center as,
\begin{eqnarray}
\rho_{ij,n}=\sum_{\tau=\tau_{\rm pu}-N_{\rm w}T}^{\tau_{\rm pu}+N_{\rm w}T}\rho_{ij}(\tau)e^{in\omega \tau} \; (n=0, \cdots,M-1),
\end{eqnarray}
by setting $N_{\rm w}$=10.

According to the Floquet theorem, the time-dependent Schr\"odinger equation for the time-periodic Hamiltonian [${\mathcal H}(\tau)={\mathcal H}(\tau+T)$] can be written in the Fourier space as~\cite{Oka_PRB09,Kitagawa_PRB10,Bukov_AP15},
\begin{eqnarray}
\sum_{m=-\infty}^{\infty}{\mathcal H}_{nm}|\Phi_{m,\lambda}\rangle=\varepsilon_{n,\lambda}
|\Phi_{n,\lambda}\rangle,
\label{eq9}
\end{eqnarray}
where ${\mathcal H}_{nm}=H_{n-m}-m\omega \delta_{n,m}$. Here $n$ and $m$ correspond to the number of absorbed or emitted photons, and $\lambda$ labels the eigenstates in each photon-number subspace. The Fourier components $H_n$ and $|\Phi_{n,\lambda}\rangle$ are given by,
\begin{eqnarray}
H_n&=&\frac{1}{T}\int^T_0{\mathcal H}(\tau)e^{in\omega \tau}d\tau, \\
|\Phi_{n,\lambda}\rangle&=&\frac{1}{T}\int^T_0|\Phi_{\lambda}(\tau)\rangle e^{in\omega \tau}d\tau,
\end{eqnarray}
where $|\Phi_{\lambda}(\tau)\rangle$ is the $\lambda$th Floquet state, which satisfies $|\Phi_{\lambda}(\tau)\rangle=|\Phi_{\lambda}(\tau+T)\rangle$. Using $\rho_{ij,n}$ and replacing ${\bm A}(\tau)$ in Eq.~(\ref{eq1}) with that for a continuous-wave light, i.e., ${\bm A}(\tau)=(E^\omega/\omega)(\cos \omega \tau,\sin \omega \tau)$, the Fourier component $H_n$ in our mean-field treatment is given by,
\begin{eqnarray}
H^{\rm MF}_n&=&\sum_{\langle i,j\rangle}t_{ij}e^{-in\theta_{ij}}[J_n(-{\mathcal A}_{ij})c^{\dagger}_{i\sigma}c_{j\sigma}
+J_n({\mathcal A}_{ij})c^{\dagger}_{j\sigma}c_{i\sigma}] \nonumber \\
&+&\sum_{i,\sigma}\Bigl(\frac{U}{2}\rho_{ii,n}+\sum_jV_{ij}\rho_{jj,n}\Bigr)n_{i\sigma} \\
&-&\frac{1}{2}\sum_{i,j,\sigma}V_{ij}\rho_{ij,n}c^{\dagger}_{j\sigma}c_{i\sigma}, \nonumber
\end{eqnarray}
where ${\mathcal A}_{ij} \equiv (E^\omega/\omega)|{\bm \delta}_{ij}|$, $\theta_{ij} \equiv \tan^{-1}(\delta^x_{ij}/\delta^y_{ij})$, and $J_n$ is the $n$th Bessel's function of the first kind. We then solve the eigenequation~(\ref{eq9}) in momentum space by restricting the number of photons to $|n|, |m|\leq 10$ to obtain eigenenergies $\varepsilon_{n,\lambda}({\bm k})$ and eigenfunctions $|\Phi_{n,\lambda}(\bm k)\rangle$.

\begin{figure}[tb]
\includegraphics[scale=0.5]{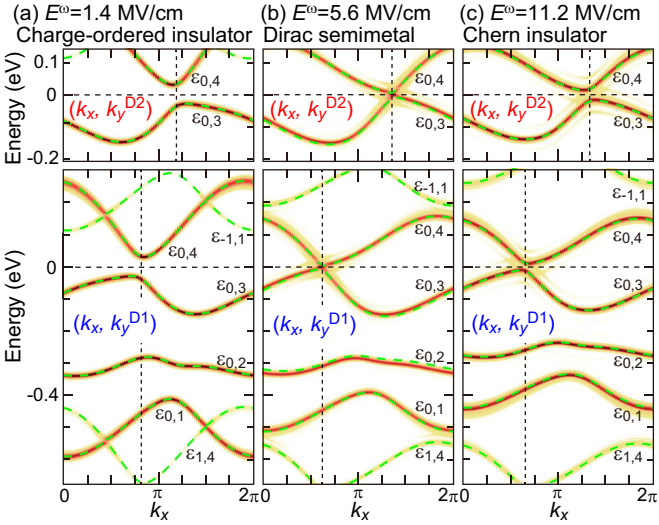}
\caption{Calculated dispersion relations of the Floquet bands $\varepsilon_{n,\lambda}(\bm k)$ along the ($k_x$, $k_y^{\rm D1}$) line (lower panels), and those along the ($k_x$, $k_y^{\rm D2}$) line around the Fermi level (upper panels) for various light amplitudes, i.e., (a) $E^\omega$=1.4 MV/cm, (b) $E^\omega$=5.6 MV/cm, and (c) $E^\omega$=11.2 MV/cm. These light amplitudes correspond to the photoinduced phases of (a) the charge-ordered insulator, (b) the Dirac semimetal, and (c) the Chern insulator as argued shortly. Here the momentum $k_y^{\rm D1}$ ($k_y^{\rm D2}$) is fixed at a constant value such that the line runs over the Dirac point located in the momentum area of $k_x<\pi$ ($k_x>\pi$).}
\label{Fig03}
\end{figure}
Figures~\ref{Fig03}(a)-(c) present the calculated dispersion relations of the Floquet bands $\varepsilon_{n,\lambda}(\bm k)$ by dashed lines. Specifically, the four bands in the zero-photon subspace $\varepsilon_{0,\lambda}$ ($\lambda$=1-4), the lowest band in the one-photon-absorbed subspace $\varepsilon_{-1,1}$, and the highest band in the one-photon-emitted subspace $\varepsilon_{1,4}$ are shown together with the transient excitation spectra $A(\bm k,\tau_{\rm pr})$ along ($k_x$, $k_y^{\rm D1}$) and ($k_x$, $k_y^{\rm D2}$) lines where the constant momentum $k_y^{\rm D1}$ ($k_y^{\rm D2}$) is selected such that the line runs over the Dirac point located in the area of $k_x<\pi$ ($k_x>\pi$).

Apparently, the Floquet bands $\varepsilon_{n,\lambda}(\bm k)$ perfectly coincide with the transient excitation spectra $A(\bm k,\tau_{\rm pr})$ over the entire $k_x$-$\varepsilon$ plane irrespective of the light amplitude $E^\omega$. This validates our approach combining the Floquet theory with the numerical simulation based on the time-dependent Schr\"odinger equation for the present interacting Dirac-fermion system. This perfect coincidence also guarantees the validity of the Berry curvatures and the Chern numbers calculated below using our theoretical formalism and the correctness of our argument on the topological nature of this system based on these physical quantities.

We also realize that the lowest ($\lambda$=1) and highest ($\lambda$=4) bands of the zero-photon subspace ($n$=0) overlap the bands of nonzero-photon subspaces of $n=+1$ and $n=-1$, respectively, when the light field is as weak as $E^\omega$=1.4 MV/cm [Fig.~\ref{Fig03}(a)]. This indicates that so-called on-resonant situation arises in the photoirradiated charge-ordered state, where the electron occupations of bands are far from those at equilibrium described by the Fermi distribution function. On the other hand, such a band overlap does not occur for stronger light fields of $E^\omega$=5.6 MV/cm [Fig.~\ref{Fig03}(b)] and $E^\omega$=11.2 MV/cm [Fig.~\ref{Fig03}(c)], indicating that the off-resonant situation with electron occupations similar to those at equilibrium is realized for the Dirac semimetal phase and the Chern insulator phase in the photodriven $\alpha$-(BEDT-TTF)$_2$I$_3$. Therefore, these two phases are well-defined and are feasible to be observed experimentally.

\begin{figure}[tb]
\includegraphics[scale=1.0]{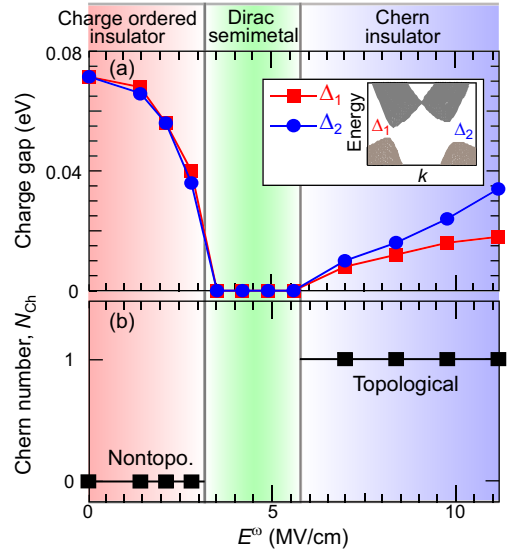}
\caption{$E^\omega$-dependence of (a) two gaps $\Delta_1$ and $\Delta_2$ at the Dirac points and (b) Chern number $N_{\rm Ch}$ in the insulator phases when $\hbar\omega$=0.7 eV.}
\label{Fig04}
\end{figure}
In Figs.~\ref{Fig04}(a) and (b), we plot the magnitudes of two gaps $\Delta_1$ and $\Delta_2$ at the Dirac points and the topological invariant called Chern number $N_{\rm Ch}$. The Chern number $N_{\rm Ch}$ is defined by the sum of the band Chern numbers $N^\lambda_{\rm Ch}$ ($\lambda$=1-4) over three bands below the Fermi level in the zero-photon subspace, when the system is an insulator, i.e., $N_{\rm Ch} \equiv \sum_{\lambda=1}^3 N^\lambda_{\rm Ch}=-N^4_{\rm Ch}$. The Chern numbers $N^\lambda_{\rm Ch}$ for respective bands are calculated from the Floquet Hamiltonian constructed above using a method in Ref.~\cite{Fukui_JPSJ05}. In the weak-field regime of $E^\omega <$3.5 MV/cm, there exist gaps, but the are suppressed with increasing $E^\omega$, indicating the photoinduced gap closing. The Chern number is zero in accord with the nontopological charge-ordered insulator in this regime. 

The Dirac semimetal with gapless Dirac-cone bands emerges in a subsequent phase. Here the Chern number is not well defined in this dynamical gapless phase. This intermediate phase emerges as a consequence of subtle competition between two effects of circularly polarized light, i.e., closing of charge gap through melting the charge order and formation of topological gap through breaking the time reversal symmetry. Surprisingly, this intermediate phase has a considerable window of $3.5 \leq E^\omega\,({\rm MV/cm}) \leq 5.6$ despite a subtle balance of this competition.

With further increasing the light amplitude $E^\omega$, gaps start opening again above $E^\omega$=5.6 MV/cm, and the Chern insulator phase with a nonzero Chern number ($N_{\rm Ch}$=1) emerges. Noticeably, the magnitudes of two gaps, $\Delta_1$ and $\Delta_2$, are inequivalent in this Chern insulator phase, which is attributable to distinct time averages of $\langle n_{\rm A}\rangle$ and $\langle n_{\rm A^\prime}\rangle$. The oscillation centers for $\langle n_{\rm A}\rangle$ and $\langle n_{\rm A^\prime}\rangle$ are not equivalent around $\tau/T\sim 250$ as seen in Fig.~\ref{Fig02}(c). This inequivalence causes the staggered site-potential of $\phi_{\rm A}$ and $\phi_{\rm A^\prime}$~\cite{Osada_JPSJ17}. This is reminiscent of an argument on the Haldane model~\cite{Haldane_PRL88} where the staggered potential at two sublattice sites on the honeycomb lattice gives rise to different gap magnitudes at the two Dirac points. A difference between our system and the Haldane model is that the topological gaps are induced dynamically by circularly polarized light in our case, while those in the Haldane model are induced by complex transfer integrals in a static system. Note that the inequivalence between $\Delta_1$ and $\Delta_2$ should appear also in the charge-ordered insulator phase, but the difference is very small because the light field is weak.


In summary, we have theoretically studied dynamical phase transitions and charge dynamics in the interacting Dirac-electron system irradiated with circularly polarized light by taking the organic compound $\alpha$-(BEDT-TTF)$_2$I$_3$ with a charge-ordered ground state. By analysing the extended Hubbard model for this compound using a combined technique of numerical simulation of the time-dependent Schr\"odinger equation and the Floquet theory, we have elucidated successive two dynamical phase transitions from the charge-ordered insulator to the gapless Dirac semimetal and finally to the Chern insulator phase as a competition between the two major effects of circularly polarized light. Our theoretical work has clearly demonstrated that driven interacting Dirac fermions host rich nonequilibrium transition phenomena and thus will necessarily stimulate the related experimental studies.

This work was partly supported by JSPS KAKENHI (Grant Nos. 20K03841 and 20H00337) and Waseda University Grant for Special Research Projects (Project Nos. 2020C-269 and 2021C-566). A part of the numerical simulations was performed at the Supercomputer Center of the Institute for Solid State Physics, the University of Tokyo.

\end{document}